\documentclass[aps,twocolumn,floatfix,floats,superscriptaddress,raggedbottom]{revtex4}
\usepackage{graphicx,latexsym,t1enc}
\usepackage{amsmath}
\begin{document}

\title{Multi-mode transport through a quantum nanowire with two embedded dots}

\author{Vidar Gudmundsson}
\affiliation{Science Institute, University of Iceland,
        Dunhaga 3, IS-107 Reykjavik, Iceland}
\author{Gudny Gudmundsdottir}
\affiliation{Science Institute, University of Iceland,
        Dunhaga 3, IS-107 Reykjavik, Iceland}
\author{Jens Hjorleifur Bardarson}
\affiliation{Science Institute, University of Iceland,
        Dunhaga 3, IS-107 Reykjavik, Iceland}
\author{Ingibjorg Magnusdottir}
\affiliation{Science Institute, University of Iceland,
        Dunhaga 3, IS-107 Reykjavik, Iceland}
\author{Chi-Shung Tang}
\affiliation{Physics Division, National Center for Theoretical
        Sciences, P.O.\ Box 2-131, Hsinchu 30013, Taiwan}
\author{Andrei Manolescu}
\affiliation{Science Institute, University of Iceland,
             Dunhaga 3, IS-107 Reykjavik, Iceland}

%

\begin{abstract}
We investigate the conductance of a quantum wire with two embedded quantum
dots using a T-matrix approach based on the Lippmann-Schwinger formalism.
The quantum dots are represented by a quantum well with Gaussian shape and
the wire is two-dimensional with parabolic confinement in the transverse
direction. In a broad wire the transport can assume a strong nonadiabatic 
character and the conductance manifests effects caused by
intertwined inter- and intra-dot processes that are identified by 
analysis of the ``nearfield'' probability distribution of the transported
electrons.   
\end{abstract}


\maketitle
\section{Introduction}
Transport through vertical double dot systems has
been investigated by many groups 
experimentally \cite{Blick96:7899,Blick98:4032,Jeong01:2221,Holleitner02:70,Rontani04:085327} 
or theoretically
\cite{Klimeck94:2316,Niu94:5130,Stafford94:3590,Aharony00:13561,Orellana02:155317,Guevara04:195335,Moldoveanu04:085303},
just to mention few. 
In most of these models the researchers use effective Hubbard-type models
or exact diagonalization for few electrons in order to describe the effects of 
higher order correlations on the electron transport.   
Other studies have focused on describing the effects of the interaction of
the geometry of the leads and the quantum dot, where commonly, the system
tracted is a dot embedded in a quantum wire \cite{Bryant91:12837,Kim99:10962}.
The focus has then been on resonances caused by the interplay of the discrete
quasi-bound states of the dot and the continuum of the wire 
\cite{Gurvitz93:10578,Noeckel94:17415}. 

Here we report our investigation on a system composed of a quasi
two-dimensional quantum wire in no external magnetic field with a parabolic 
confinement in the direction transverse to the transport.
Embedded in the quantum wire we have two identical quantum dots whose centers
are separated by the distance $2d$, but the effective size of the smoothly 
Gaussian shaped dots is on the order of $d/8$. We shall consider a narrow
wire chosen such that the depth of the dot is a bit larger than the
energy separation of the unperturbed subbands of the wire, and a broad wire where
the subband separation is only $1/10$ of the dot depth. We consider two
different dot sizes, but in both cases the effective size of the dot
does not broaden the wire. We are thus considering a wire with embedded
dots in between the limit where the dots are similar to an impurity represented 
by attractive $\delta$-functions, and the case where the dots can be
considered of the same width or wider than the wire.   
We assume the electrons to be incident in some of the lowest lying energy subbands of the
wire, but due to the geometry of the system we have to include several more subbands  
in the calculation in order to describe the transport
correctly.    
\section{Models}
We consider a multi-mode transport of electrons along the $z$-direction  
through a two-dimensional quantum wire defined by a parabolic confinement 
in the $x$-direction with the characteristic frequency $E_0=\hbar\Omega_0$.   
The electrons incident from the left ($z\rightarrow -\infty$) impinge on a
scattering potential composed of two Gaussian wells located at $z=\pm d$
\begin{equation}
      V_{sc}=V_0\left\{ e^{-\beta ((z-d)^2 + x^2) } + 
                         e^{-\beta ((z+d)^2 + x^2) } \right\} .
\label{eq:Vsc}
\end{equation}
The scattering modes are expanded in the transverse modes $\chi_m$
\begin{equation}
      \psi_{nE}^+(\mathbf{r}) = \sum_m \varphi_{mE}^n(z)\chi_m(x),
\label{eq:psi}
\end{equation}
that are the eigenmodes of the parabolic confinement of the 
quantum wire. An incident electron with energy 
$E=\hbar^2k^2_n(E)/(2m)+\epsilon_n$, where $\epsilon_n = \hbar\Omega_0(n+1/2)$
can propagate in mode $n$ if $\epsilon_n < E$ otherwise the state is
evanescent. We use the coupled multi-mode Lippmann-Schwinger 
equations \cite{Bagwell90:10354,Cattapan03:903}
to establish a set of coupled integral equations for the $T$-matrix 
\cite{Bardarson04:01} which in turn is
used to calculate the transmission amplitude that within the Landauer
formalism leads to the linear response conductance at
vanishing temperature \cite{fisher81:6851,BILP85}
\begin{equation}
      G  = \frac{2e^2}{h}\text{Tr}[t^\dagger t].
\label{eq:Landauer}
\end{equation}
The $T$-matrix can also be used to construct the probability density of the
scattering states through their wave functions \cite{Bardarson04:01}
\begin{equation}
  \begin{split}
  \phi_{mE}^n(z)
    = \phi_{mE}^{n0}(z) + \frac{m}{\pi\hbar^2}
    \int_{-\infty}^{+\infty} 
    dp \frac{\sqrt{|p|}e^{ipz}}{k_m^2-p^2}T_{mn}(p,k_n). 
  \end{split}
  \label{eq:phi_T}
\end{equation}
All matrix elements have been calculated analytically and
special care has been taken to include enough modes in the numerical 
calculations to reach convergent solutions. The singular part of
the integration for the scattering states (\ref{eq:phi_T}) and the
$T$-matrix has been completed analytically \cite{Bardarson04:01}.    

The mirror symmetry of the scattering potential (\ref{eq:Vsc}) along the
center, $x=0$, of the wire results in a finite matrix elements only between
transverse modes of the same parity.   
\section{Results}
We shall consider a broad ($E_0=\hbar\Omega_0=1$ meV) or a 
narrow ($E_0=\hbar\Omega_0=6$ meV) quantum wire with two identical embedded quantum dots
located at $z=\pm d$. The dots are small, with
$\beta=0.01$ nm$^{-2}$, or larger with $\beta=0.003$ nm$^{-2}$ in Eq.\ (\ref{eq:Vsc}).  
The depth of the dots or wells is $V_0=-10$ meV, and we assume GaAs
parameters yielding the effective Bohr radius $a_0 = 9.79$ nm, and 
the Rydberg $Ry=5.92$ meV. Clearly, we can expect the transport to vary
strongly with the width of the wire since in the case of the broad wire 
several subbands or modes can be coupled by the scattering potential.
In the narrow wire the characteristic length $a_w = \sqrt{\hbar /(m\Omega_0)} = 
13.8$ nm, in the broad one $a_w = 33.7$ nm.

\subsection{Narrow wire} 
\begin{figure}[tbhp!]
      \includegraphics[width=0.45\textwidth]{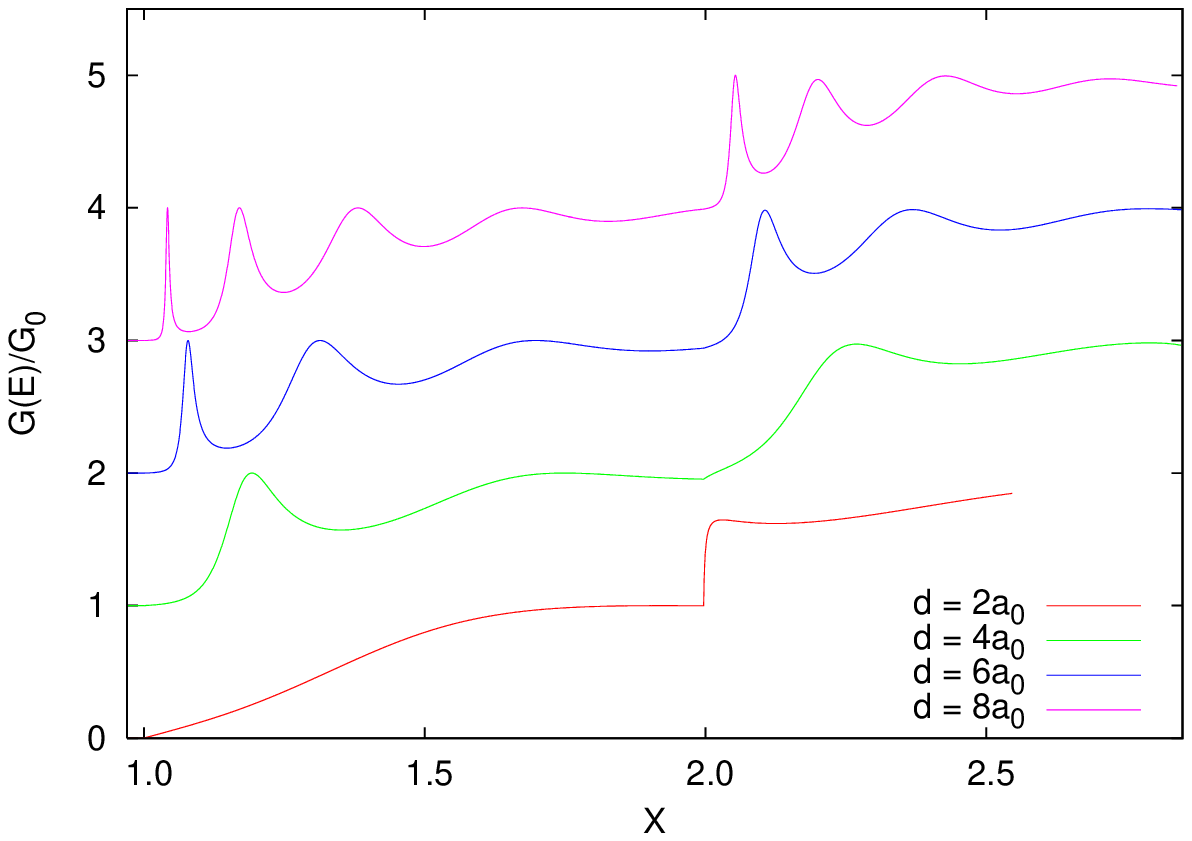}\\
      \includegraphics[width=0.45\textwidth]{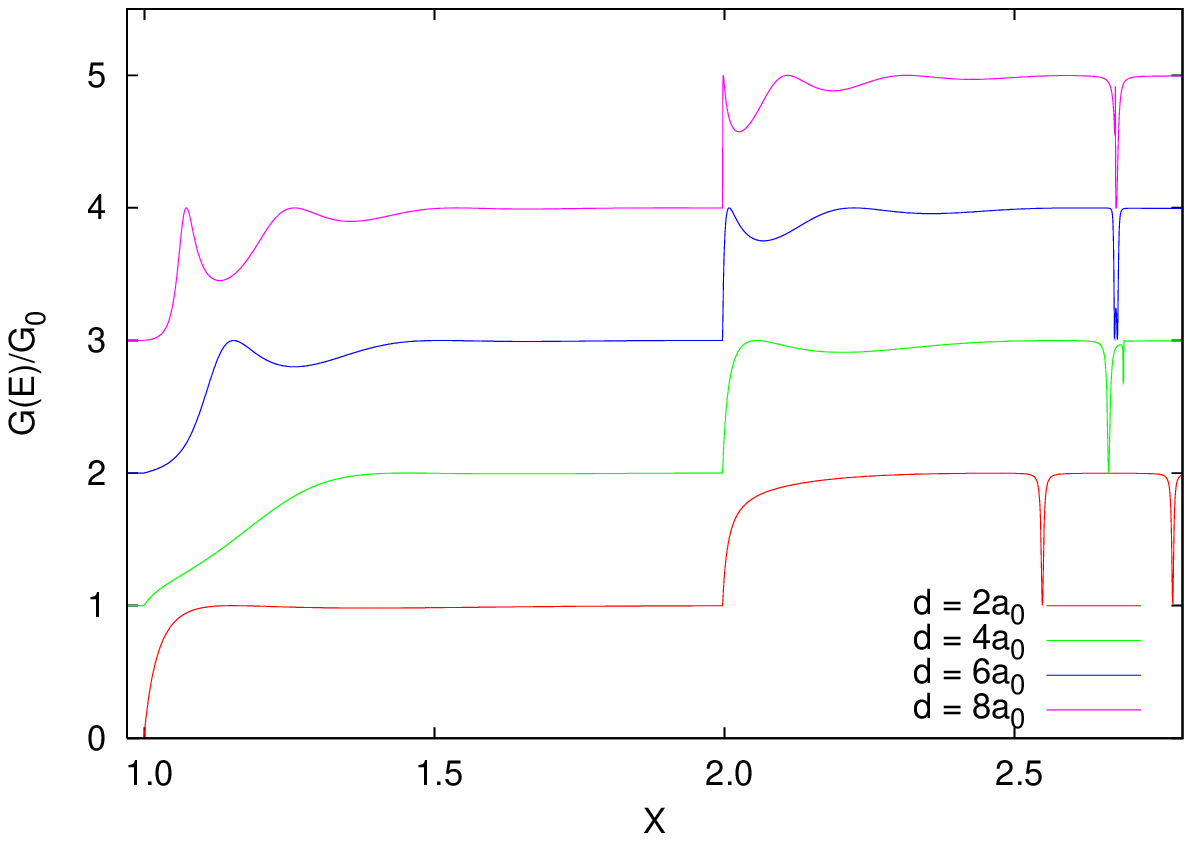}
      \caption{The conductance in units of $G_0=2e^2/h$
               of a narrow parabolic wire with embedded small 
               dots $\beta a_w^2=1.90$ (upper panel), and large dots
               $\beta a_w^2=0.571$ (lower panel) for different
               displacement $z=\pm d$
               from the center of the wire $z=0$ as
               function of $X=E/E_0+1/2$.
               $E_0=\hbar\Omega_0=6.0$meV, $V_0=-10$ meV,
               $a_w=13.8$ nm, and for GaAs $a_0=9.79$ nm.}
\label{fig:G_MjoirBreidir}
\end{figure}

\begin{figure}[tbhp!]
      \includegraphics[width=0.45\textwidth]{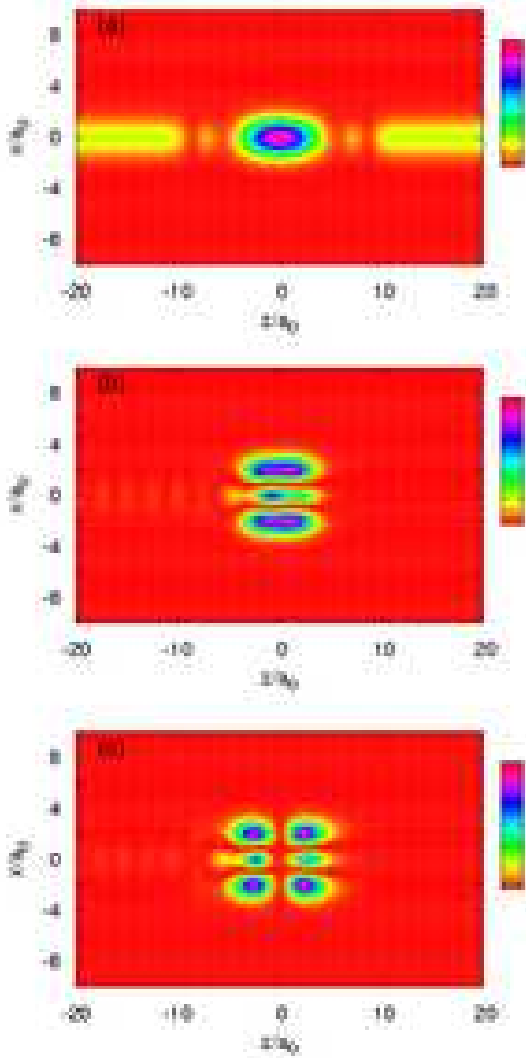}
      \caption{The probability density of the scattering state $\psi_{E}(x,y)$
               in the narrow quantum wire in the presence of
               two embedded large dots.
               The incident energy of the $n=0$ mode corresponds to $X=1.08$ and $d=
               8a_0$ (a), 
               $X=2.58$ and $d= 2a_0$ (b), and 
               $X=2.80$ and $d= 2a_0$ (c),
               corresponding to the peak and the two dips,
               respectively, in the conductance in the lower panel of 
               Fig.\ \ref{fig:G_MjoirBreidir}.
               $E_0=\hbar\Omega_0=6.0$ meV, $V_0=-10$ meV,
               $a_w=13.8$ nm, $\beta a_w^2=0.571$.}
\label{fig:W_MjoirBreidir}
\end{figure}

The conductance of the narrow wire is shown in Fig.\ \ref{fig:G_MjoirBreidir}
for both types of dots as function of the parameter $X=E/E_0+1/2$, whose
integral part numbers the propagating subbands 
participating in the transport at a particular energy $E$. 
The figure indicates that the main role of the dots
in the narrow wire is to define a semitransparent cavity in between them.
In the case of the small dots this cavity has a well defined length and 
well known geometrical resonances are seen in the conductance as the
separation, $d$, of the dots is increased. When the dots are larger
(lower panel) the higher order resonances vanish and one or two dips occur
in the conductance, characteristic of an attractive scattering potential.  
For the smallest separation of the dots, when they partially overlap we see
two dips. In Fig.\ \ref{fig:W_MjoirBreidir} we present the probability density
for three relevant cases in order to better understand the microscopic
processes in the system. In Fig.\ \ref{fig:W_MjoirBreidir}(a) we see the
probability density for the large dots corresponding to the lowest
resonance in the lower panel of Fig.\ \ref{fig:G_MjoirBreidir}, when
$d = 8a_0$. As expected the main probability density is located in the
cavity between the dots. Interestingly, due to the finite size of the dots 
we also see a small density at the location of each dot. Here the electrons
enter the system in the lowest mode $n=0$ and exit the system in the same
mode, this together with the total transmission leads to a constant probability 
density in the wire away from the scattering potential. 
In Fig.\ \ref{fig:W_MjoirBreidir}(b) and (c) we see the probability density 
at the dips for the large dots as $d = 2a_0$. The two states causing the dips are the
symmetric and the anti-symmetric quasi-bound state the latter one occurring
at a bit higher energy as expected. The structure of the density in either
case reminds us that due to the high symmetry of the dot potential,  
Eq.\ (\ref{eq:Vsc}), the evanescent state causing the reflection of the
$n=0$ mode is in the third subband with $n=2$ and is thus broader than the
resonant state in Fig.\ \ref{fig:W_MjoirBreidir}(a). Due to the high probability
density of the electrons in the evanescent state we do barely see the incoming
and the reflected density on the left side of the scattering potential
on the chosen color scale.     

\subsection{Broad wire}
\begin{figure}[tbhp!]
      \includegraphics[width=0.45\textwidth]{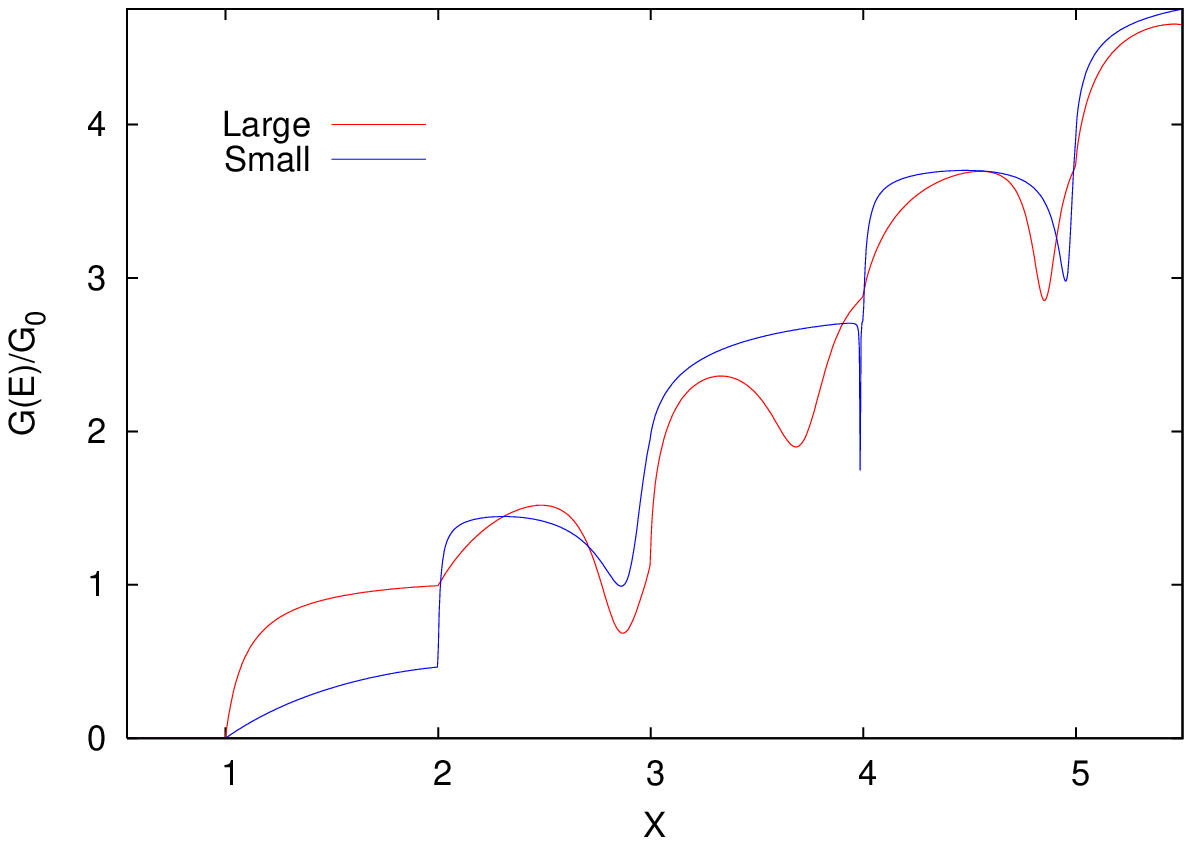}\\
      \includegraphics[width=0.45\textwidth]{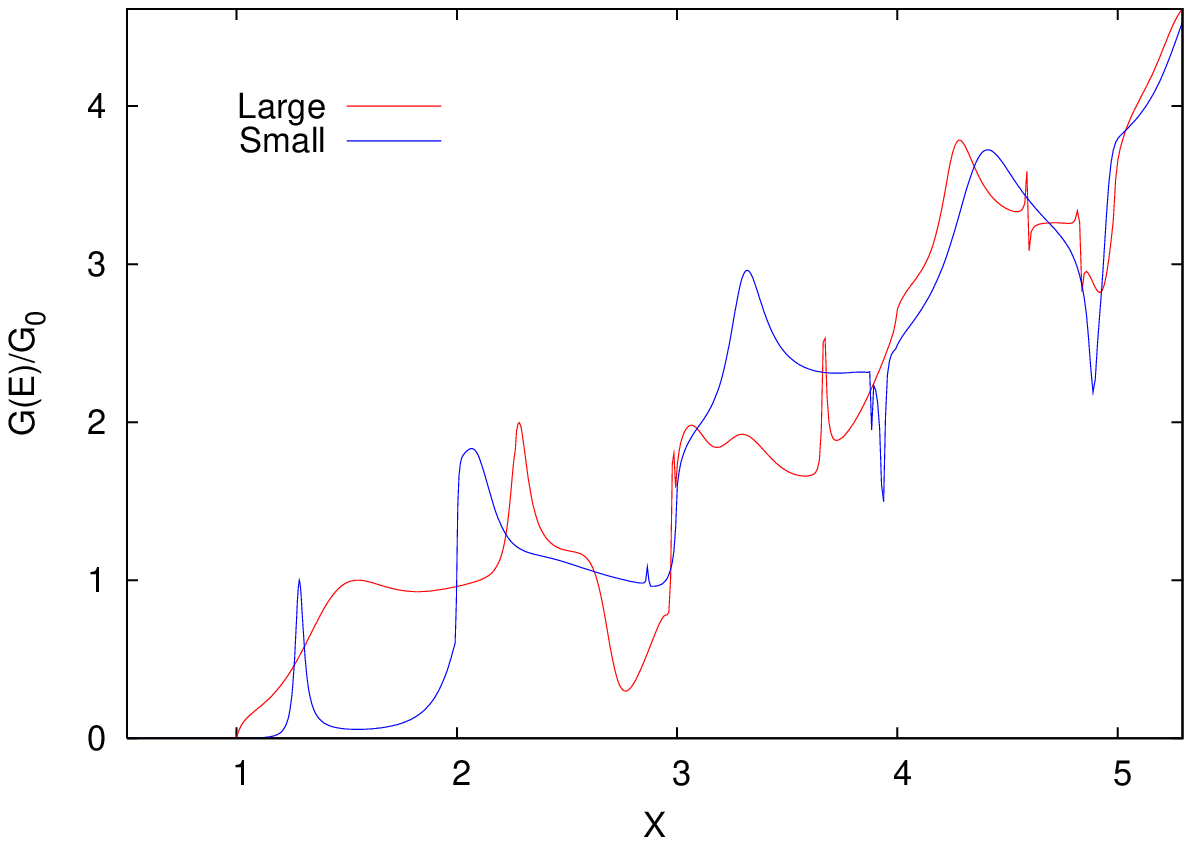}
      \caption{The conductance in units of $G_0=2e^2/h$
               of a broad wire with one embedded  
               dot at the center $z=0$, (upper panel),
               and two dots at $z=\pm 8a_0$, (lower panel)
               as function of $X=E/E_0+1/2$.  
               $E_0=\hbar\Omega_0=1.0$ meV, $V_0=-10$ meV,
               $a_w=33.7$ nm, and for GaAs $a_0=9.79$ nm.
               A small dot is characterized by $\beta a_w^2=11.37$
               and a large one by $\beta a_w^2=3.41$.}
\label{fig:G_GB}
\end{figure}

In order to prepare the analyses of the conduction of a broad quantum wire
($E_0=1$ meV) with two embedded dots we first show in the upper panel of 
Fig.\ \ref{fig:G_GB} the conduction in the broad wire with only one quantum 
dot embedded, large or small. The sharp dip seen at the end of the second 
conduction step in a narrow wire is turned into a broad minimum in the broad 
wire due to the stronger coupling between the subbands caused by the dot 
in the broad wire.  

In the lower panel of Fig.\ \ref{fig:G_GB} we display the conductance of
the broad wire with the two embedded dots at $d = 8a_0$. 
The presence of the two well separated dots adds considerably to the 
fine structure of the conductance compared to the results for one dot.
Moreover, this structure involves the coupling of many subbands and
thus requires the inclusion of approximately 16 of them in the numerical 
calculation in order to reach well converged results.     

\begin{figure}[tbhp!]
      \includegraphics[width=0.45\textwidth]{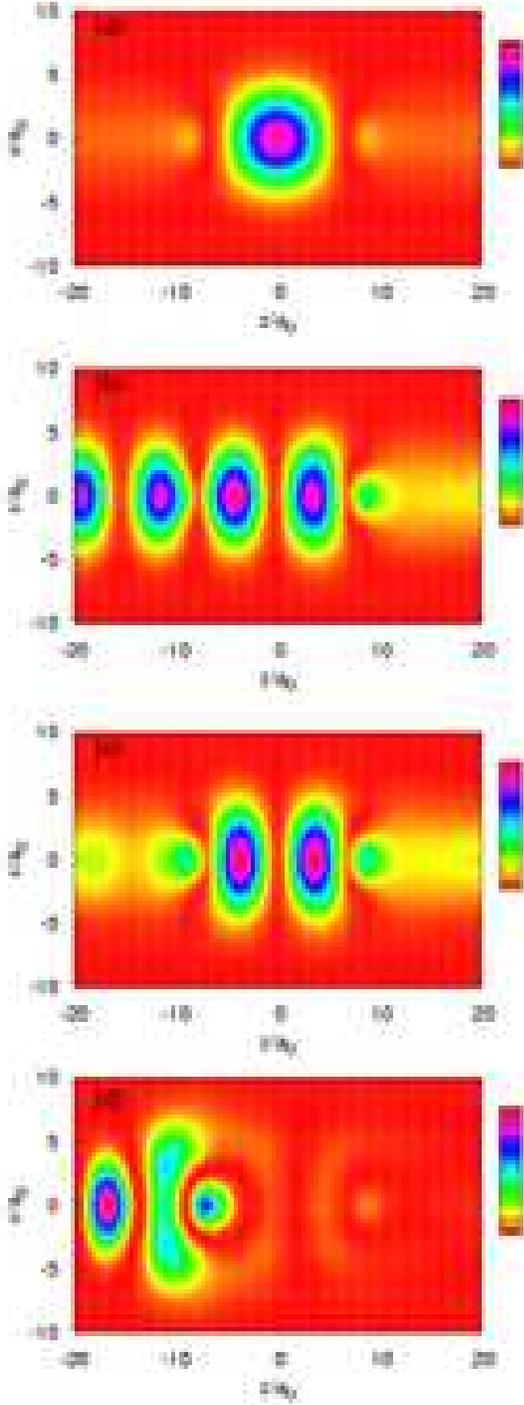}
      \caption{The probability density of the scattering state $\psi_{E}(x,y)$
               in the broad quantum wire in the presence of
               two small embedded dots centered at $z=\pm 8a_0$. 
               The incident energy and modes correspond to:
               (a) $X=1.284$ and $n=0$, (b) $X=1.98$ and $n=0$,
               (c) $X=2.07$ and $n=0$, and (d) $X=2.76$ and $n=0$.  
               $E_0=\hbar\Omega_0=1.0$ meV, $V_0=-10$ meV,
               $a_w=33.7$ nm, and $\beta a_w^2=11.37$.} 
\label{fig:Fig_WG_01}
\end{figure}

First, we start with the electron probability density shown in Fig.\
\ref{fig:Fig_WG_01} to gain insight into the multitude of processes
taking place in the case of the small dots. Similarly, as in the case of the 
narrow wire we have a transmission resonance at $X=1.284$ caused by a
state located in the ``cavity'' between the dots shown in 
Fig.\ \ref{fig:Fig_WG_01}(a). We also see small probability maxima at the
dots themselves. The second transmission resonance at $X=2.07$ is caused
by the next mode in the inter-dot cavity and is shown in Fig.\
\ref{fig:Fig_WG_01}(c), but it is also interesting to check the probability
density at a bit lower energy $X=1.98$, see \ref{fig:Fig_WG_01}(b).
Here there is considerable reflection and as the Figure shows this can be 
interpreted as the electrons being reflected by the second dot, the first
dot simply loses its effects in a minimum in the interference between
the incoming and reflected wave. Thus, by increasing the energy to $X=2.76$
we observe a reflection of the first dot in \ref{fig:Fig_WG_01}(d).
The valley structure in $G$ around $X$ = $2.76$ corresponds
to the electrons making inter-subband transitions from $n$=$0$ to
the subband threshold of the third subband ($n$=$2$) forming a quasi-bound 
state. The wide valley structure implies a
short life time of such a state.    

\begin{figure}[tbhp!]
      \includegraphics[width=0.45\textwidth]{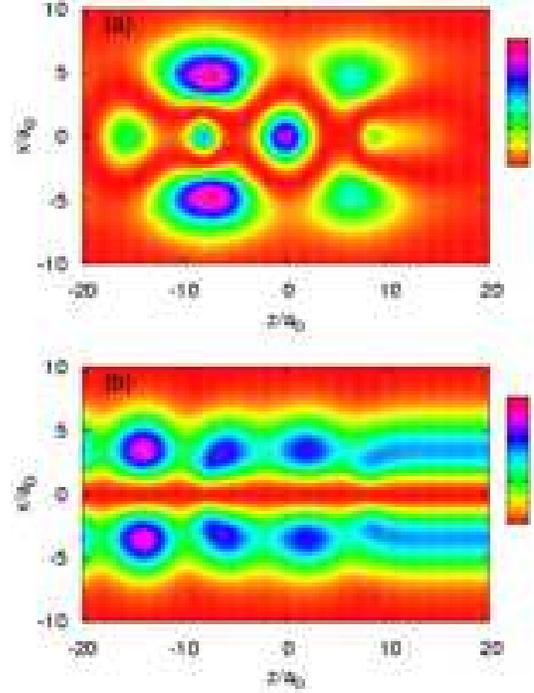}
      \caption{The probability density of the scattering state $\psi_{E}(x,y)$
               in the broad quantum wire in the presence of
               two small embedded dots centered at $z=\pm 8a_0$. 
               The incident energy and modes correspond to:
               (a) $X=2.86$ and $n=0$, (b) $X=2.86$ and $n=1$.
               $E_0=\hbar\Omega_0=1.0$ meV, $V_0=-10$ meV,
               $a_w=33.7$ nm, and $\beta a_w^2=11.37$.}
\label{fig:Fig_WG_02}
\end{figure}

At $X=2.86$ we see a narrow resonance in the
conduction. The corresponding probability density in Fig.\ \ref{fig:Fig_WG_02}
has the typical structure of an evanescent state in the third subband for
the incoming $n=0$ state and a normal conducting mode for the $n=1$ instate.
The evanescent state clearly belongs mainly to the first dot, 
but is also fairly extended into the inter-dot cavity. 

\begin{figure}[tbhp!]
      \includegraphics[width=0.45\textwidth]{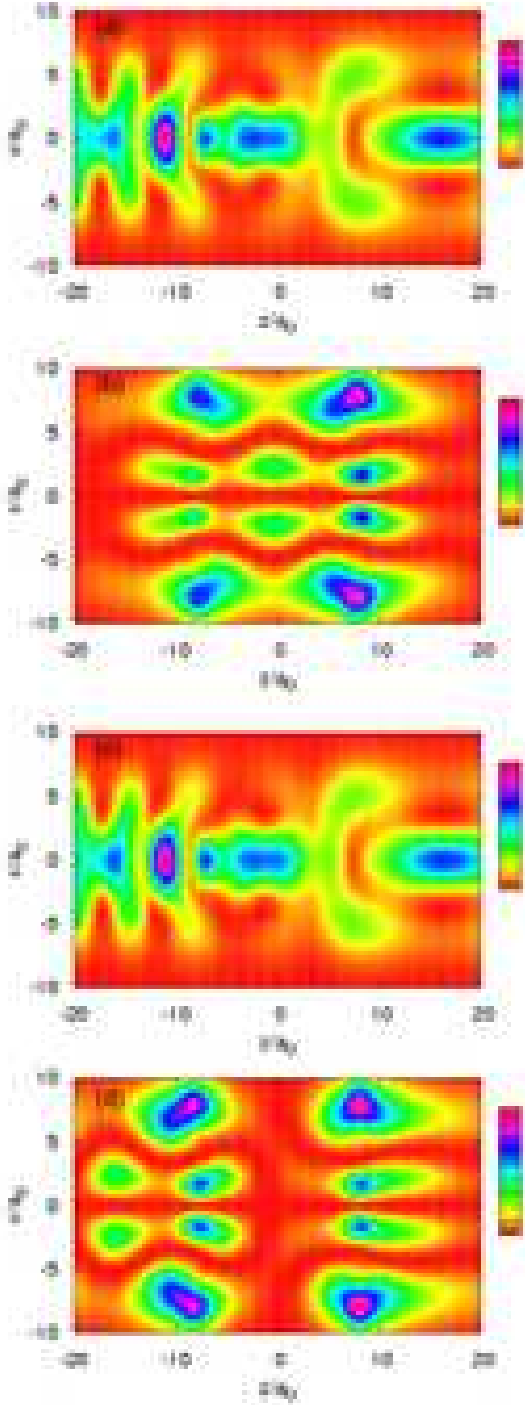}
      \caption{The probability density of the scattering state $\psi_{E}(x,y)$
               in the broad quantum wire in the presence of
               two small embedded dots centered at $z=\pm 8a_0$. 
               The incident energy and modes correspond to:
               (a) $X=3.88$ and $n=0$, (b) $X=3.88$ and $n=1$,
               (c) $X=3.94$ and $n=0$, and (d) $X=3.94$ and $n=1$.  
               $E_0=\hbar\Omega_0=1.0$ meV, $V_0=-10$ meV,
               $a_w=33.7$ nm, and $\beta a_w^2=11.37$.}
\label{fig:Fig_WG_03}
\end{figure}

To finish the observation of the case of the small dots we examine the
electronic probability density for the two narrow dips found at 
$X=3.88$, and 3.94. These are presented in Fig.\ \ref{fig:Fig_WG_03}
for the two lowest incoming modes $n=0$ and 1.  
Here the first and the third mode conduct partially, as can be confirmed 
for the $n=0$ mode in Fig.\ \ref{fig:Fig_WG_03}(a) and (c). The two
$n=0$ modes are almost identical, but the second incoming mode $n=1$ is
reflected due to an interaction with an evanescent mode in the fourth $n=3$
subband. The evanescent modes in Fig.\ \ref{fig:Fig_WG_03}(b) and (d) show
the symmetry of the fourth band and are the distinct two lowest symmetric and
antisymmetric quasi-bound states of the dot system.

\begin{figure}[tbhp!]
      \includegraphics[width=0.45\textwidth]{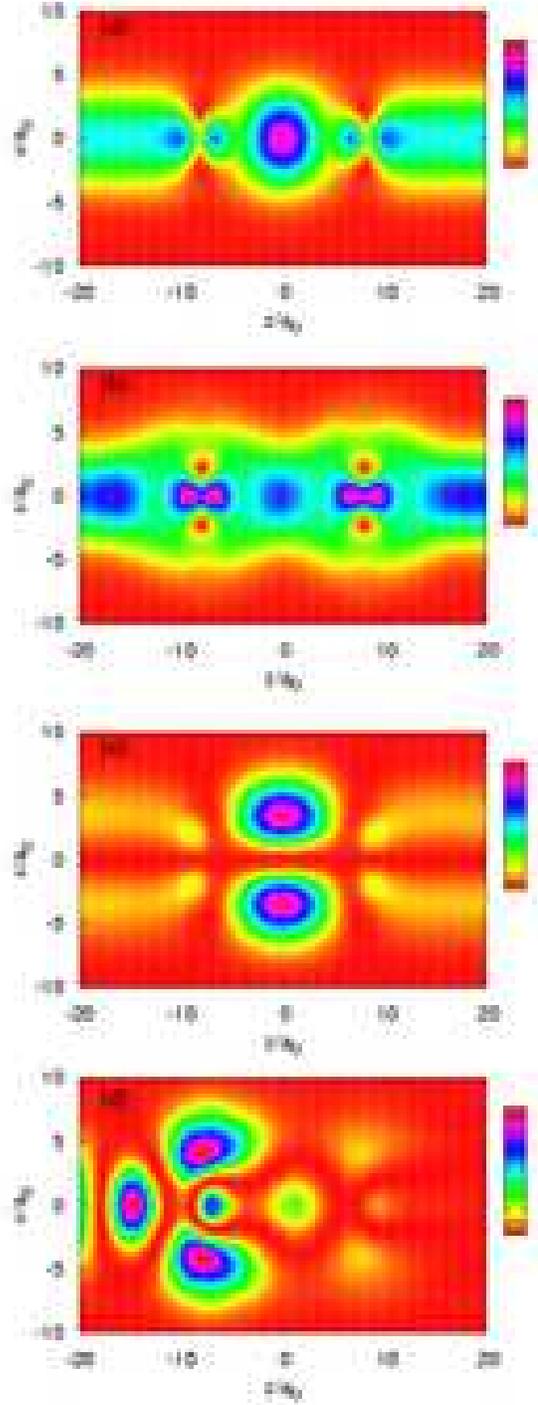}
      \caption{The probability density of the scattering state $\psi_{E}(x,y)$
               in the broad quantum wire in the presence of
               two large embedded dots centered at $z=\pm 8a_0$. 
               The incident energy and modes correspond to:
               (a) $X=1.55$ and $n=0$, (b) $X=2.28$ and $n=0$,
               (c) $X=2.28$ and $n=1$, (d) $X=2.77$ and $n=0$ 
               $E_0=\hbar\Omega_0=1.0$ meV, $V_0=-10$ meV,
               $a_w=33.7$ nm, and $\beta a_w^2=3.41$.}
\label{fig:Fig_WB_01}
\end{figure}

The system with larger dots allows for a richer mixture of intra
and inter-dot processes or states. The lowest transmission resonance 
is now a broad feature with a corresponding electronic probability 
density that is a clear superposition of a state in the inter-dot cavity
and states located at each dot, see Fig.\ \ref{fig:Fig_WB_01}(a).
The next transmission resonance at $X=2.28$ is narrow and for the
incoming $n=0$ mode is composed of a state with a high proportion of
a $p_z$-type atomic orbital or higher on each dot as the probability in
Fig.\ \ref{fig:Fig_WB_01}(b) shows. The $n=1$ in-mode on the
other hand is basically a $p_y$-type state in the inter-dot cavity and
to lesser degree $p_y$-states at each dot, see Fig.\ \ref{fig:Fig_WB_01}(c).
In the conductance minimum at $X=2.77$ we observe the familiar
evanescent mode (Fig.\ \ref{fig:Fig_WB_01}(d)) in the third subband
reflecting a large portion of the incoming wave. 

\begin{figure}[tbhp!]
      \includegraphics[width=0.45\textwidth]{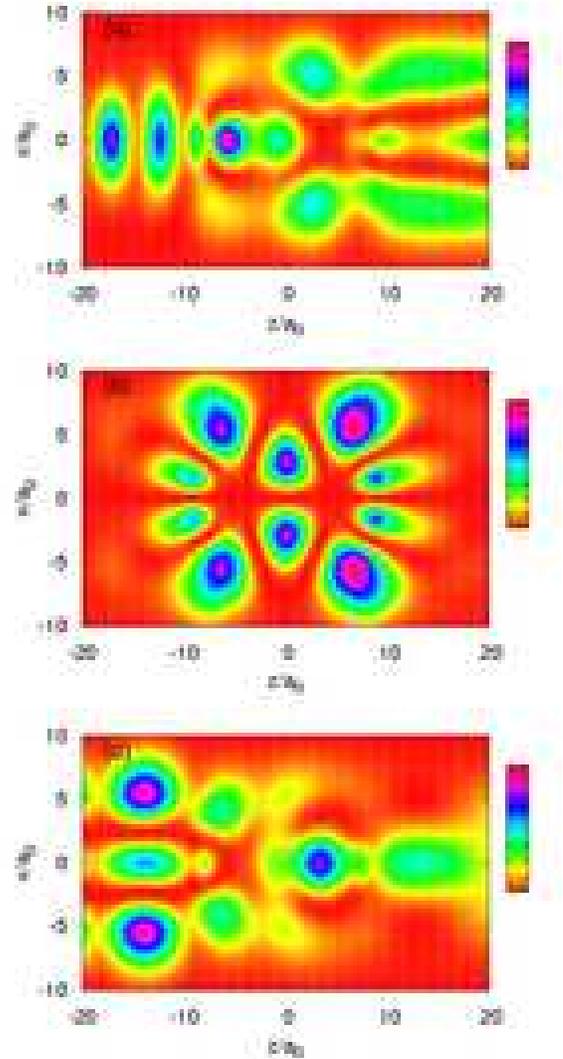}
      \caption{The probability density of the scattering state $\psi_{E}(x,y)$
               in the broad quantum wire in the presence of
               two large embedded dots centered at $z=\pm 8a_0$. 
               The incident energy and modes correspond to:
               (a) $X=3.67$ and $n=0$, (b) $X=3.67$ and $n=1$,
               and (c) $X=3.67$ and $n=2$ 
               $E_0=\hbar\Omega_0=1.0$ meV, $V_0=-10$ meV,
               $a_w=33.7$ nm, and $\beta a_w^2=3.41$.}
\label{fig:Fig_WB_02}
\end{figure}

At $X=3.67$ the conductance displays a narrow resonance that is caused
by an evanescent state in the fourth subband interacting with the second
incoming mode as can be verified by the probability density shown
in Fig.\ \ref{fig:Fig_WB_02}(b). Here is also curious to note that 
the scattering potential mixes up the first and the third incoming modes;
Fig.\ \ref{fig:Fig_WB_02}(a) shows the incoming $n=0$ state leaving
the scattering region in the $n=2$ mode, and the opposite process is 
seen happening in Fig.\ \ref{fig:Fig_WB_02}(c). In the first case there is
a considerable probability for the electron in the first dot, and in the
second case this is reversed. 
Here, one should remember that even though we talk about large dots in this
case the effective size of the dots is comparable to the width of the
incoming state and thus the third mode is quite broader than the dot,
and the broadness of the wire leads to large matrix elements between the
the first and the third mode of the wire.
This mixing of incoming states, a character of nonadiabatic transport, 
or crosstalk between the channels or modes, 
is not limited to the peak at $X=3.67$. At the broader peak at $X=4.3$
the same also happens for the odd modes $n=2$ and $n=4$.

As expected, for higher energy the conductance of the broad wire becomes
similar for the small and the larger dots except for the Fano resonances
that represent quasi-bound states present in the energy continuum that
depend on the shape of the pertinent dots.     
\section{Summary}
We have explored the transport through a quantum wire with two embedded
quantum dots that are not broader then the wire. We find in the transport
an interplay between intra- and inter-dot scattering processes.  
In particular, we find for the larger dots in a broad wire drastic
expamles of nonadiabatic transport \cite{Bryant91:12837}.
Bryant explores the transport through a hard-wall quantum wire with a quantum dot
defined by a tapered center and separated from the wire region by two
rectangular barriers \cite{Bryant91:12837}. He finds the mode-mixing effects
to be more important when the transmission occurs by resonant
tunneling and when the tapering is not quite smooth. Here we observe strong
mode-mixing in ballistic transport when the energy subbands of the wire
are closely spaced on the scale of the depth of the quantum dots and the
size of the dots results in large matrix elements between the various
transverse modes. Here is important that the symmetry of the centered
dots establishes selection rules for the matrix elements that are thus of
very variable magnitude. Also, even though one quantum dot of Gaussian 
shape can be considered smooth, then it is more difficult to measure
the smoothness of the system of two smooth dots very well separated.  

The methods employed to calculate the conductance have 
been applied to wires with parabolic (here) and hard wall
confinement \cite{Bardarson04:01} in the transverse direction 
and are applicable to general scattering potentials as long as care 
is taken in selecting a sufficient number of input modes.

\begin{acknowledgments}
      The research was partly funded by the Research 
      and Instruments Funds of the Icelandic State,
      and the Research Fund of the University of Iceland.
      C.S.T.\ acknowledges computational facilities supported
      by the National Center for High-performance Computing in Taiwan.
\end{acknowledgments}

%
%
\bibliographystyle{apsrev}

\begin{thebibliography}{21}
\expandafter\ifx\csname natexlab\endcsname\relax\def\natexlab#1{#1}\fi
\expandafter\ifx\csname bibnamefont\endcsname\relax
  \def\bibnamefont#1{#1}\fi
\expandafter\ifx\csname bibfnamefont\endcsname\relax
  \def\bibfnamefont#1{#1}\fi
\expandafter\ifx\csname citenamefont\endcsname\relax
  \def\citenamefont#1{#1}\fi
\expandafter\ifx\csname url\endcsname\relax
  \def\url#1{\texttt{#1}}\fi
\expandafter\ifx\csname urlprefix\endcsname\relax\def\urlprefix{URL }\fi
\providecommand{\bibinfo}[2]{#2}
\providecommand{\eprint}[2][]{\url{#2}}

\bibitem[{\citenamefont{Blick et~al.}(1996)\citenamefont{Blick, Haug, Weis,
  Pfannkuche, v.~Klitzing, and Eberl}}]{Blick96:7899}
\bibinfo{author}{\bibfnamefont{R.~H.} \bibnamefont{Blick}},
  \bibinfo{author}{\bibfnamefont{R.~J.} \bibnamefont{Haug}},
  \bibinfo{author}{\bibfnamefont{J.}~\bibnamefont{Weis}},
  \bibinfo{author}{\bibfnamefont{D.}~\bibnamefont{Pfannkuche}},
  \bibinfo{author}{\bibfnamefont{K.}~\bibnamefont{v.~Klitzing}},
  \bibnamefont{and} \bibinfo{author}{\bibfnamefont{K.}~\bibnamefont{Eberl}},
  \bibinfo{journal}{Phys. Rev. B} \textbf{\bibinfo{volume}{53}},
  \bibinfo{pages}{7899} (\bibinfo{year}{1996}).

\bibitem[{\citenamefont{Blick et~al.}(1998)\citenamefont{Blick, Pfannkuche,
  Haug, Klitzing, and Eberl}}]{Blick98:4032}
\bibinfo{author}{\bibfnamefont{R.~H.} \bibnamefont{Blick}},
  \bibinfo{author}{\bibfnamefont{D.}~\bibnamefont{Pfannkuche}},
  \bibinfo{author}{\bibfnamefont{R.~J.} \bibnamefont{Haug}},
  \bibinfo{author}{\bibfnamefont{K.}~\bibnamefont{Klitzing}}, \bibnamefont{and}
  \bibinfo{author}{\bibfnamefont{K.}~\bibnamefont{Eberl}},
  \bibinfo{journal}{Phys. Rev. Lett.} \textbf{\bibinfo{volume}{80}},
  \bibinfo{pages}{4032} (\bibinfo{year}{1998}).

\bibitem[{\citenamefont{Jeong et~al.}(2001)\citenamefont{Jeong, Chang, and
  Melloch}}]{Jeong01:2221}
\bibinfo{author}{\bibfnamefont{H.}~\bibnamefont{Jeong}},
  \bibinfo{author}{\bibfnamefont{A.~M.} \bibnamefont{Chang}}, \bibnamefont{and}
  \bibinfo{author}{\bibfnamefont{M.~R.} \bibnamefont{Melloch}},
  \bibinfo{journal}{Science} \textbf{\bibinfo{volume}{293}},
  \bibinfo{pages}{2221} (\bibinfo{year}{2001}).

\bibitem[{\citenamefont{Holleitner et~al.}(2002)\citenamefont{Holleitner,
  Blick, H{\"u}ttel, Eberl, and Kotthaus}}]{Holleitner02:70}
\bibinfo{author}{\bibfnamefont{A.~W.} \bibnamefont{Holleitner}},
  \bibinfo{author}{\bibfnamefont{R.~H.} \bibnamefont{Blick}},
  \bibinfo{author}{\bibfnamefont{A.~K.} \bibnamefont{H{\"u}ttel}},
  \bibinfo{author}{\bibfnamefont{K.}~\bibnamefont{Eberl}}, \bibnamefont{and}
  \bibinfo{author}{\bibfnamefont{J.~P.} \bibnamefont{Kotthaus}},
  \bibinfo{journal}{Science} \textbf{\bibinfo{volume}{297}},
  \bibinfo{pages}{70} (\bibinfo{year}{2002}).

\bibitem[{\citenamefont{Rontani et~al.}(2004)\citenamefont{Rontani, Amaha,
  Muraki, Manghi, Molinari, Tarucha, and Austing}}]{Rontani04:085327}
\bibinfo{author}{\bibfnamefont{M.}~\bibnamefont{Rontani}},
  \bibinfo{author}{\bibfnamefont{S.}~\bibnamefont{Amaha}},
  \bibinfo{author}{\bibfnamefont{K.}~\bibnamefont{Muraki}},
  \bibinfo{author}{\bibfnamefont{F.}~\bibnamefont{Manghi}},
  \bibinfo{author}{\bibfnamefont{E.}~\bibnamefont{Molinari}},
  \bibinfo{author}{\bibfnamefont{S.}~\bibnamefont{Tarucha}}, \bibnamefont{and}
  \bibinfo{author}{\bibfnamefont{D.~G.} \bibnamefont{Austing}},
  \bibinfo{journal}{Phys. Rev. B} \textbf{\bibinfo{volume}{69}},
  \bibinfo{pages}{085327} (\bibinfo{year}{2004}).

\bibitem[{\citenamefont{Klimeck et~al.}(1994)\citenamefont{Klimeck, Chen, and
  Datta}}]{Klimeck94:2316}
\bibinfo{author}{\bibfnamefont{G.}~\bibnamefont{Klimeck}},
  \bibinfo{author}{\bibfnamefont{G.}~\bibnamefont{Chen}}, \bibnamefont{and}
  \bibinfo{author}{\bibfnamefont{S.}~\bibnamefont{Datta}},
  \bibinfo{journal}{Phys. Rev. B} \textbf{\bibinfo{volume}{50}},
  \bibinfo{pages}{2316} (\bibinfo{year}{1994}).

\bibitem[{\citenamefont{Niu et~al.}(1994)\citenamefont{Niu, jun Liu, and han
  Lin}}]{Niu94:5130}
\bibinfo{author}{\bibfnamefont{C.}~\bibnamefont{Niu}},
  \bibinfo{author}{\bibfnamefont{L.}-\bibnamefont{jun Liu}}, \bibnamefont{and}
  \bibinfo{author}{\bibfnamefont{T.}-\bibnamefont{han Lin}},
  \bibinfo{journal}{Phys. Rev. B} \textbf{\bibinfo{volume}{51}},
  \bibinfo{pages}{5130} (\bibinfo{year}{1994}).

\bibitem[{\citenamefont{Stafford and Sarma}(1994)}]{Stafford94:3590}
\bibinfo{author}{\bibfnamefont{C.~A.} \bibnamefont{Stafford}} \bibnamefont{and}
  \bibinfo{author}{\bibfnamefont{S.~D.} \bibnamefont{Sarma}},
  \bibinfo{journal}{Phys. Rev. Lett.} \textbf{\bibinfo{volume}{72}},
  \bibinfo{pages}{3590} (\bibinfo{year}{1994}).

\bibitem[{\citenamefont{Aharony et~al.}(2000)\citenamefont{Aharony,
  Entin-Wohlmann, Imry, and Levinson}}]{Aharony00:13561}
\bibinfo{author}{\bibfnamefont{A.}~\bibnamefont{Aharony}},
  \bibinfo{author}{\bibfnamefont{O.}~\bibnamefont{Entin-Wohlmann}},
  \bibinfo{author}{\bibfnamefont{Y.}~\bibnamefont{Imry}}, \bibnamefont{and}
  \bibinfo{author}{\bibfnamefont{Y.}~\bibnamefont{Levinson}},
  \bibinfo{journal}{Phys. Rev. B} \textbf{\bibinfo{volume}{62}},
  \bibinfo{pages}{13561} (\bibinfo{year}{2000}).

\bibitem[{\citenamefont{Orellana et~al.}(2002)\citenamefont{Orellana, Lara, and
  Anda}}]{Orellana02:155317}
\bibinfo{author}{\bibfnamefont{P.~A.} \bibnamefont{Orellana}},
  \bibinfo{author}{\bibfnamefont{G.~A.} \bibnamefont{Lara}}, \bibnamefont{and}
  \bibinfo{author}{\bibfnamefont{E.~V.} \bibnamefont{Anda}},
  \bibinfo{journal}{Phys. Rev. B} \textbf{\bibinfo{volume}{65}},
  \bibinfo{pages}{155317} (\bibinfo{year}{2002}).

\bibitem[{\citenamefont{de~Guevara et~al.}(2004)\citenamefont{de~Guevara,
  Claro, and Orellana}}]{Guevara04:195335}
\bibinfo{author}{\bibfnamefont{M.~L.~L.} \bibnamefont{de~Guevara}},
  \bibinfo{author}{\bibfnamefont{F.}~\bibnamefont{Claro}}, \bibnamefont{and}
  \bibinfo{author}{\bibfnamefont{P.~A.} \bibnamefont{Orellana}},
  \bibinfo{journal}{Phys. Rev. B} \textbf{\bibinfo{volume}{67}},
  \bibinfo{pages}{195335} (\bibinfo{year}{2004}).

\bibitem[{\citenamefont{Moldoveanu et~al.}(2004)\citenamefont{Moldoveanu,
  Aldea, and Tanatar}}]{Moldoveanu04:085303}
\bibinfo{author}{\bibfnamefont{V.}~\bibnamefont{Moldoveanu}},
  \bibinfo{author}{\bibfnamefont{A.}~\bibnamefont{Aldea}}, \bibnamefont{and}
  \bibinfo{author}{\bibfnamefont{B.}~\bibnamefont{Tanatar}},
  \bibinfo{journal}{Phys. Rev. B} \textbf{\bibinfo{volume}{70}},
  \bibinfo{pages}{085303} (\bibinfo{year}{2004}).

\bibitem[{\citenamefont{Bryant}(1991)}]{Bryant91:12837}
\bibinfo{author}{\bibfnamefont{G.~W.} \bibnamefont{Bryant}},
  \bibinfo{journal}{Phys. Rev. B} \textbf{\bibinfo{volume}{44}},
  \bibinfo{pages}{12837} (\bibinfo{year}{1991}).

\bibitem[{\citenamefont{Kim et~al.}(1999)\citenamefont{Kim, Satanin, Joe, and
  Cosby}}]{Kim99:10962}
\bibinfo{author}{\bibfnamefont{C.~S.} \bibnamefont{Kim}},
  \bibinfo{author}{\bibfnamefont{A.~M.} \bibnamefont{Satanin}},
  \bibinfo{author}{\bibfnamefont{Y.~S.} \bibnamefont{Joe}}, \bibnamefont{and}
  \bibinfo{author}{\bibfnamefont{R.~M.} \bibnamefont{Cosby}},
  \bibinfo{journal}{Phys. Rev. B} \textbf{\bibinfo{volume}{60}},
  \bibinfo{pages}{10962} (\bibinfo{year}{1999}).

\bibitem[{\citenamefont{Gurvitz and Levinson}(1993)}]{Gurvitz93:10578}
\bibinfo{author}{\bibfnamefont{S.~A.} \bibnamefont{Gurvitz}} \bibnamefont{and}
  \bibinfo{author}{\bibfnamefont{Y.~B.} \bibnamefont{Levinson}},
  \bibinfo{journal}{Phys. Rev. B} \textbf{\bibinfo{volume}{47}},
  \bibinfo{pages}{10578} (\bibinfo{year}{1993}).

\bibitem[{\citenamefont{N{\"o}ckel and Stone}(1994)}]{Noeckel94:17415}
\bibinfo{author}{\bibfnamefont{J.~U.} \bibnamefont{N{\"o}ckel}}
  \bibnamefont{and} \bibinfo{author}{\bibfnamefont{A.~D.} \bibnamefont{Stone}},
  \bibinfo{journal}{Phys. Rev. B} \textbf{\bibinfo{volume}{50}},
  \bibinfo{pages}{17415} (\bibinfo{year}{1994}).

\bibitem[{\citenamefont{Bagwell}(1990)}]{Bagwell90:10354}
\bibinfo{author}{\bibfnamefont{P.~F.} \bibnamefont{Bagwell}},
  \bibinfo{journal}{Phys. Rev. B} \textbf{\bibinfo{volume}{41}},
  \bibinfo{pages}{10354} (\bibinfo{year}{1990}).

\bibitem[{\citenamefont{Cattapan and Maglione}(2003)}]{Cattapan03:903}
\bibinfo{author}{\bibfnamefont{G.}~\bibnamefont{Cattapan}} \bibnamefont{and}
  \bibinfo{author}{\bibfnamefont{E.}~\bibnamefont{Maglione}},
  \bibinfo{journal}{Am. J. Phys.} \textbf{\bibinfo{volume}{71}},
  \bibinfo{pages}{903} (\bibinfo{year}{2003}).

\bibitem[{\citenamefont{Bardarson et~al.}(2004)\citenamefont{Bardarson,
  Magnusdottir, Gudmundsdottir, Tang, Manolescu, and
  Gudmundsson}}]{Bardarson04:01}
\bibinfo{author}{\bibfnamefont{J.~H.} \bibnamefont{Bardarson}},
  \bibinfo{author}{\bibfnamefont{I.}~\bibnamefont{Magnusdottir}},
  \bibinfo{author}{\bibfnamefont{G.}~\bibnamefont{Gudmundsdottir}},
  \bibinfo{author}{\bibfnamefont{C.-S.} \bibnamefont{Tang}},
  \bibinfo{author}{\bibfnamefont{A.}~\bibnamefont{Manolescu}},
  \bibnamefont{and}
  \bibinfo{author}{\bibfnamefont{V.}~\bibnamefont{Gudmundsson}},
  \bibinfo{journal}{(cond-mat/0408435)}  (\bibinfo{year}{2004}).

\bibitem[{\citenamefont{Buttiker et~al.}(1985)\citenamefont{Buttiker, Imry,
  Landauer, and Pinhas}}]{BILP85}
\bibinfo{author}{\bibfnamefont{M.}~\bibnamefont{Buttiker}},
  \bibinfo{author}{\bibfnamefont{Y.}~\bibnamefont{Imry}},
  \bibinfo{author}{\bibfnamefont{R.}~\bibnamefont{Landauer}}, \bibnamefont{and}
  \bibinfo{author}{\bibfnamefont{S.}~\bibnamefont{Pinhas}},
  \bibinfo{journal}{Phys. Rev. B} \textbf{\bibinfo{volume}{31}},
  \bibinfo{pages}{6207} (\bibinfo{year}{1985}).

\bibitem[{\citenamefont{Fisher and Lee}(1981)}]{fisher81:6851}
\bibinfo{author}{\bibfnamefont{D.~S.} \bibnamefont{Fisher}} \bibnamefont{and}
  \bibinfo{author}{\bibfnamefont{P.~A.} \bibnamefont{Lee}},
  \bibinfo{journal}{Phys. Rev. B} \textbf{\bibinfo{volume}{23}},
  \bibinfo{pages}{6851} (\bibinfo{year}{1981}).

\end{thebibliography}

%
%
%
\end{document}